\documentclass[fleqn,10pt]{SelfArx} % Document font size and equations flushed left
\definecolor{color1}{RGB}{0,0,90} % Color of the article title and sections
\definecolor{color2}{RGB}{0,20,20} % Color of the boxes behind the abstract and headings
\usepackage{hyperref} % Required for hyperlinks
\hypersetup{hidelinks,colorlinks,breaklinks=true,urlcolor=color2,citecolor=color1,linkcolor=color1,bookmarksopen=false,pdftitle={Title},pdfauthor={Author}}

\usepackage{mathptmx}       % selects Times Roman as basic font
\usepackage{helvet}         % selects Helvetica as sans-serif font
\usepackage{courier}        % selects Courier as typewriter font
\usepackage{type1cm}        % activate if the above 3 fonts are
\usepackage{makeidx}         % allows index generation
\usepackage{graphicx}        % standard LaTeX graphics tool
\graphicspath{{Figures/}}
\usepackage{multicol}        % used for the two-column index
\usepackage[bottom]{footmisc}% places footnotes at page bottom
\usepackage{amssymb}

\usepackage{amsthm}
 
\theoremstyle{definition}
\newtheorem{definition}{Definition}[section]
\graphicspath{ {./Figures/} }
\DeclareMathAlphabet{\mathcal}{OMS}{cmsy}{m}{n}
\SetMathAlphabet{\mathcal}{bold}{OMS}{cmsy}{b}{n}% see the list of further useful packages
\makeindex             % used for the subject index
\JournalInfo{This is a preprint of a chapter from the book \textit{Temporal Network Theory}. \\Cite as:} % Journal information
\Archive{\textit{R. Cazabet, G. Rossetti (2019). Challenges in Community Discovery on Temporal Networks. In Holme and Saramaki (Editors) Temporal Network Theory. Springer- Nature, New York. 2019.}} % Additional notes (e.g. copyright, DOI, review/research article)

\PaperTitle{Challenges in Community Discovery on Temporal Networks} % Article title
\Authors{Remy Cazabet\textsuperscript{1}*, Giulio Rossetti\textsuperscript{2}} % Authors
\affiliation{\textsuperscript{1}Univ Lyon, UCBL, CNRS, LIRIS UMR 5205, F-69621, Lyon, France, } % Author affiliation
\affiliation{\textsuperscript{2}Knowledge Discovery and Data Mining Lab, ISTI-CNR, Pisa, Italy } % Author affiliation
 % Keywords - if you don't want any simply remove all the text between the curly brackets

\Abstract{Community discovery is one of the most studied problems in network science. In recent years, many works have focused on discovering communities in temporal networks, thus identifying dynamic communities. Interestingly, dynamic communities are not mere sequences of static ones; new challenges arise from their dynamic nature. In this chapter, we will discuss some of these challenges and recent propositions to tackle them. We will, among other topics, discuss on the question of community events in gradually evolving networks, on the notion of identity through change, on dynamic communities in link streams, on the smoothness of dynamic communities, and on the different types of complexity of algorithms for their discovery. %and of the notion of community life-cycle.
}

\begin{document}
\flushbottom % Makes all text pages the same height

\maketitle % Print the title and abstract box

\tableofcontents % Print the contents section

\section*{Introduction}
The modular nature of networks is one of the most studied aspects of network science. In most real-world networks, a mesoscale organization exists, with nodes belonging to one or several \textit{modules} or \textit{clusters}\cite{newman2006modularity}: think of groups in social networks (groups of friends, families, organizations, countries, etc.), or biological networks such as brain networks \cite{meunier2010modular}.  The term \textit{community} is commonly used in the network science literature to describe a \textit{set} of nodes that are grouped for \textit{topological} reasons (e.g., they are strongly connected together and more weakly connected to the rest of the network. Other topological criteria exist, such as having a high internal clustering, similar connection patterns, etc. See Section \ref{goodCommunities} for more on this topic). The literature on the topic is large and diverse, not only on the topic of automatic community discovery but also on community evaluation, analysis, or even generation of networks with realistic community structure. In the last ten years, many works have focused on adapting those problems to temporal networks\cite{rossetti2018community}. In this chapter, we present an overview of the active topics of research on dynamic communities. For each of these topics, when relevant, we highlight some current challenges.

The chapter is organized into five parts. In the first one, we discuss the definition of dynamic \textit{clusters} in temporal networks, and how to represent them. In the second section, we concentrate on the specificity of dynamic communities, in particular focusing on \textit{smoothness}, \textit{identity} and \textit{algorithmic complexity}. Section 3 focuses on the differences between communities in different types of dynamic networks such as link streams or snapshot sequences. In section 4, we discuss the evaluation of dynamic communities, using internal and external evaluation --requiring appropriate synthetic benchmarks. Finally, in section 5, we briefly introduce existing tools to work with dynamic communities.

\section{Representing dynamic communities}
\label{definition}
The first question to answer when dealing with communities is: \textit{what is a good community?}
There is no universal consensus on this topic in the literature; thus, in this article, we adopt a definition as large as possible: 
\begin{definition}[Community]
A (static) community  in a graph $G=(V,E)$ is i) a \textit{cluster} (i.e., a \textit{set}) of nodes $C \subseteq V$ ii) having \textit{relevant} topological characteristics as defined by a community detection algorithm.
\end{definition}
The second part of this definition will be discussed in section \ref{goodCommunities}, and is concerned by the question of the \textit{quality} of a set of nodes as a community, based on a topological criterion. On the contrary, this section discusses the transposition of the first part of this definition to temporal networks, i.e., the definition of \textit{dynamic node clusters} themselves, independently of any quality criteria. We use the term cluster in its data analysis meaning, i.e., clusters are groups of items defined such as those items are more similar (in some sense) to each other than to those in other groups (clusters).

To define dynamic clusters, we first need to define what is a temporal network. This question will be discussed in detail in section \ref{differentTypes}. For now, let's adopt a generic definition provided in \cite{Latapy2018}, representing in an abstract way any type of temporal network:
\begin{definition}[Temporal Network]
A temporal network, or stream graph, is defined as $S=(T,V,W,E)$, with $V$ a set of nodes, $T$ a set of time instants (continuous or discrete), $W \subseteq T \times V$, and $E \subseteq T \times V \otimes V$.
\end{definition}

\subsection{Fixed membership cluster in temporal networks}
The first possible transposition of static clusters to temporal networks is to consider memberships as fixed:
\begin{definition}[Fixed Membership Cluster]
A fixed membership cluster is defined on a temporal network $S=(T,V,W,E)$ as a cluster of nodes $C \subseteq V$
%, such as this cluster is considered a \textit{relevant community} when considering the whole temporal network.
\end{definition}
In fixed membership clusters, nodes cannot change community along time.
Communities identified using this definition in a temporal network are usually considered relevant when the clustering they induce would be considered relevant according to a static definition of communities (e.g., modularity) in most times $t$ of the temporal network. 
Those communities are different from static ones found in the aggregated graph in that they take into account the \textit{temporal order} of edges. Note that in some algorithms such as stochastic block models, in which communities are defined not only by sets of nodes but also by properties of relations between communities, those properties might evolve, while membership themselves stay unchanged (e.g., \cite{matias2015estimation}). This approach can also be combined with \textit{change point detection} to find periods of the graph with \textit{stable} community structures \cite{Peel204}.

\subsection{Evolving-membership clusters in temporal networks}
In this second transposition of the definition of cluster, nodes can change membership along time. Note that, for methods based on \textit{crisp} communities, each node must belong to one (and only one) community at each step, while less constrained methods allow having nodes not belonging to any community (conversely, belonging to several communities), in some or all steps.

\begin{definition}[Evolving-Membership Cluster]
An evolving-membership cluster is defined on a temporal network $S=(T,V,W,E)$ as a cluster $C = \{(t,v), (t,v) \subseteq W\}$
\end{definition}
Dynamic communities using this type of clusters are usually considered relevant when i) the clusters it defines at each $t$ would be considered relevant according to a static definition of communities (e.g., modularity) at each step $t$, and ii) the clusters it defines at time $t$ are relatively similar to those belonging to the same dynamic cluster at $t-1$ and $t+1$. This is related to the notion of dynamic community \textit{smoothness} discussed in section \ref{smoothness}.

\subsubsection*{Persistent-labels formalism}
The usual way to implement this definition is by using what we call the \textit{persistent labels} formalism: community identifiers --labels-- are associated with some nodes over some periods. There is, therefore, no notion of being an \textit{ancestor/descendent} of another community: two nodes can either share a common label, and therefore be part of the same dynamic community, or not. This representation is the most widespread, used for instance in \cite{Mucha2010,Falkowski2006}. 

\subsection{Evolving-membership clusters with events}

\begin{figure*}[!]\centering
\centering
\includegraphics[width=0.7\linewidth]{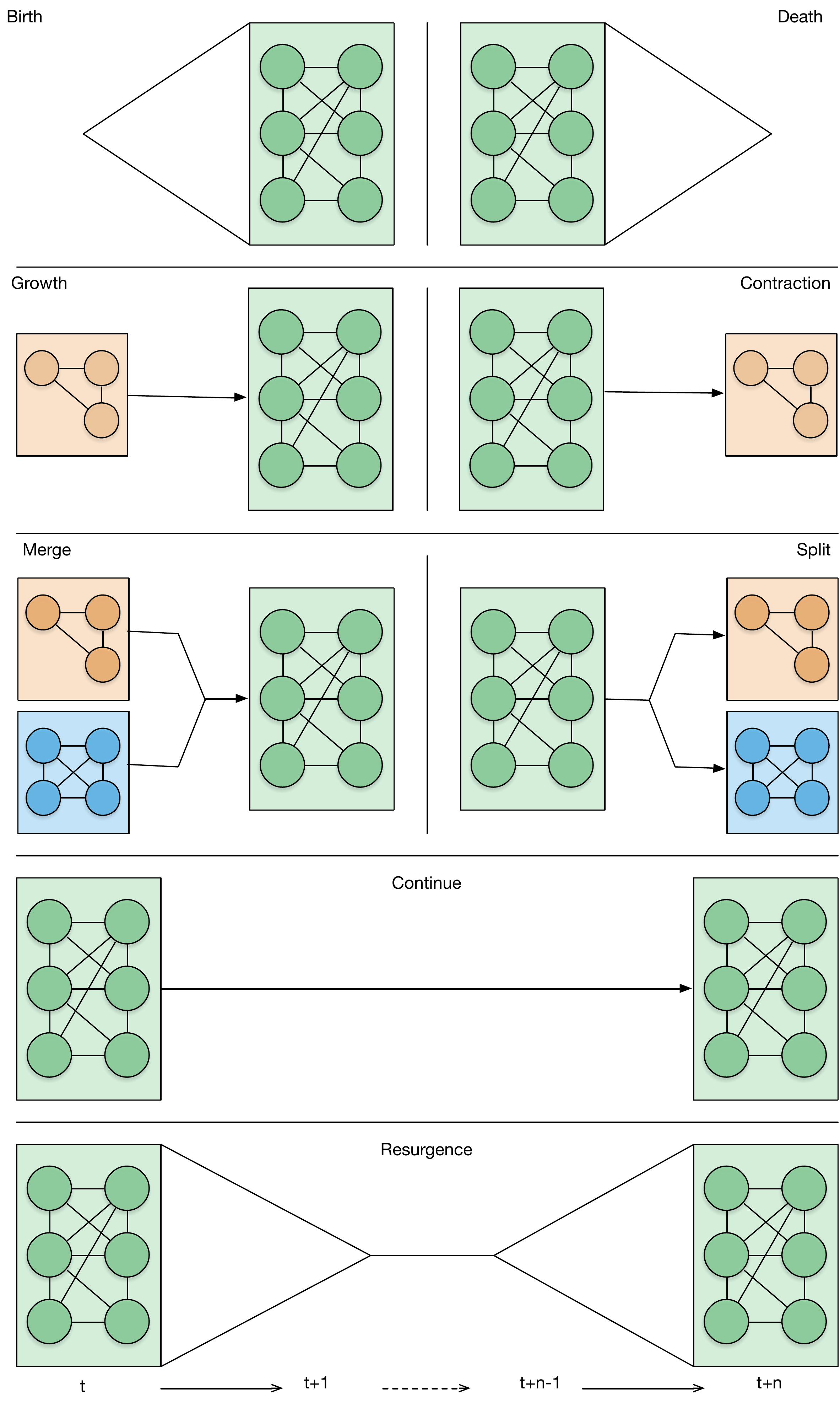}
\caption{Different types of community events}
\label{fig:events}
\end{figure*}

One of the most interesting features of dynamic communities is that they can undergo \textit{events}. Their first formal categorization was introduced in \cite{Palla2007}, which listed six of them (birth, death, growth, contraction, merge, and split).
A seventh operation, continue, is sometimes added to these. In \cite{cazabet2014}, an eighth operation was proposed (resurgence).
These events, illustrated in Figure \ref{fig:events}, are the following:
\begin{itemize}
    \item \textbf{Birth}: The first appearance of a new community composed of any number of nodes.
    \item \textbf{Death}: The vanishing of a community: all nodes belonging to the vanished community lose this membership.
    \item \textbf{Growth}: New nodes increase the size of a community.
    \item \textbf{Contraction}: Some nodes are lost by a community, thus reducing its size.
    \item \textbf{Merge}: Two communities or more merge into a single one.
    \item \textbf{Split}: A community, as a consequence of node/edge vanishing, splits into two or more components.
    \item \textbf{Continue}: A community remains unchanged in consecutive time steps.
    \item \textbf{Resurgence}: A community vanishes for a period, then comes back without perturbations as if it has never stopped existing. This event can be seen as a fake death-birth pair involving the same node set over a lagged period (e.g., seasonal behaviors).
\end{itemize}
Not all operations are necessarily handled by a generic Dynamic Community Detection algorithm.

Let's consider a situation in which two communities merge at time $t$. Using the \textit{persistent-labels} formalism introduced previously, this event can be represented in two ways: either both clusters disappear at time $t$ and a new one --the result of the merge-- is created, or one of the clusters becomes the merged one from time $t$, and the other --considered absorbed-- disappear. In both cases, important information is lost. A third definition of evolving membership can be used to solve this problem:
\begin{definition}[Evolving-membership clusters with events]
Evolving membership cluster with events are defined on a temporal network $S=(T,V,W,E)$ as a set of \textit{fixed-membership Cluster} defined at each time $t$ (or as a set of \textit{evolving-membership clusters}), and a set of community events $F$. Those events can involve several clusters (merge, split), or a single one (birth, death, shrink, etc.)
\end{definition}

\subsubsection*{Event-graph formalism}
In practice, most algorithms that do detect events record them in an ad-hoc manner (e.g., the same event can be recorded as: "a split event occurred to community $c_1$ at time $t$, yielding communities $c_1$ and $c_2$" or "community $c_2$ was born at time $t$, spawn from $c_1$"). Different representations might even be semantically different.
A few works, notably \cite{greene2010tracking}, have used an alternative way to represent dynamic communities and events, using what we call here an \textit{event-graph}. We define it as follows:

\begin{definition}[Event Graph]
An event graph is an oriented graph representing dynamic communities of the temporal network $S=(T,V,W,E)$, in which each node corresponds to a pair $\langle C ,t\rangle $, with $C \subseteq V,t  \subseteq T$, and each directed edge represents a relation of continuity between two communities, directed from the earlier to the latter.
\end{definition}
Using this representation, some events can be characterized using nodes in/out degrees:
\begin{itemize}
    \item \textbf{In-degree=0} represents new-born communities
    \item \textbf{In-degree$\geq$2} represents merge events
    \item \textbf{Out-degree=0} represents death events
    \item \textbf{Out-degree$\geq$2} represents split events
\end{itemize}

Events represented by an event graph can be much more complex than simple merge/split, since, for instance, a node-community can have multiple out-going links towards node-community having themselves multiple incoming ones.
\\ \ \\
Both representations, \textit{event-graph} and \textit{persistent labels}, have advantages and drawbacks. The former can represent any event or relation between different communities at different times, while the later can identify which community is \textit{the same as} which other one in a different time.

\subsection{Community Life-Cycle}
\label{lifecycle}

%One of the main reason to adapt the Community Discovery problem to dynamic network is to provide a coherent way to study and characterize different states and events involving clusters of network entities. 
%Pursuing such goal, several authors make  ``community tracking" the final application for the DCD algorithm they propose, often agreeing on the set of simple actions that involve entities of a dynamic network: node/edge appearance and vanishing. 
%Dynamic communities, when they are defined as evolving ones, are characterized by two local and atomic operations: node can generate perturbations of the network topology able to affect the results produced by CD algorithms. 
%As a consequence of this set of actions, given a community $C$ observed at different moments in time, it is mandatory to characterize the transformations it undergoes. 
% Dynamic communities with evolving memberships can be characterized by sequences of transformations, that can be of several nature.

% Not all operations are necessarily handled by a generic DCD algorithm. 
% Even though the abstract semantics of such transformations are clear, different works propose to handle them differently.
% Among them, merge, split, and resurgence are often defined with the support of ad hoc similarity and thresholding functions, as well as community transformation strategies. 

%Despite the differences between approaches when defining topological transformations, the 
Identified events allow to describe for each cluster the \textit{life-cycle} of its corresponding community:

\begin{definition}[Community Life-Cycle]
Given a community $C$, its life-cycle (which univocally identifies $C$'s complete evolution history) is composed of the directed acyclic graph (DAG) such that (i) the roots are birth events of $C$, and of its potential predecessors if $C$ has been
implicated in merge events; (ii) the leafs are death events, corresponding to deaths of $C$ and of its successors, if $C$ has been implicated in split events; and (iii) the central nodes are the remaining actions of $C$, its successors, and predecessors. The edges of the tree represent transitions between subsequent actions in $C$ life.
\end{definition}

\subsubsection*{Challenges}
Usual events such as birth, merge or shrink were designed to describe a few steps of evolution in the context of snapshot graphs, but are not well suited to describe complex dynamics in networks studied at a fine temporal granularity. In real scenarios, communities are susceptible to evolve gradually. A $shrink$ event might corresponds to different scenarios, such as a node switching to another community, a node leaving the system (disappearing), or the community spouting a newborn community composed of a subset of its nodes --and maybe, of other nodes. The usual representation with only labels, even with the addition of some simple events, might be too limited to represent the full range of possible community life-cycle.
Defining a complete framework to represent formally complex community evolution scenarios therefore represents a challenge for researchers in the field.

\section{Detecting dynamic communities}
\label{goodCommunities}
Defining what are good communities in networks is already a challenge in itself. Community discovery is often used as an umbrella term for several related problems, not sharing the same formal objective. It stems from earlier, well-defined problems, in particular, \textit{graph partitioning}, which consists, for a graph and given properties of a partition (number and size of clusters), to find affiliations of nodes minimizing the number of inter-cluster edges. This problem is well-defined, in that its objective can be expressed unequivocally in mathematical terms, and has no trivial solution. But having to provide the number and size of communities was considered too constraining when working with real networks having unknown properties. New methods were therefore introduced, based on ideas such as the modularity \cite{newman2004finding}, compression of random walks \cite{rosvall2008maps}, stochastic block models (SBM) and minimal description length (MDL) \cite{peixoto2014hierarchical}, intrinsic properties of communities, and so on. While some of them --e.g., modularity-- are based on the same principle of keeping (exceptionally) low the number of inter-community edges, other techniques are searching for completely different things, such as methods based on the Stochastic Block Model framework, in which \textit{blocks} are groups of nodes sharing a similar pattern of connections with nodes belonging to other groups. Furthermore, communities are often categorized in overlapping --one node can belong to several communities-- and non-overlapping (crisp) clustering methods. In this chapter, we make abstraction of those differences: each algorithm has a definition of what are \textit{good} static communities, and what we focus on are challenges introduced when going from static to dynamic ones, in particular the notions of temporal smoothness, of identity preservation, and finally the problem of scalability of existing algorithms.

\subsection{Different approaches of temporal Smoothness}
\label{smoothness}
% The wide range of heterogeneous choices that can be taken while describing dynamic communities specific problems, as well as the increasing number of algorithms that have been designed to address them, inevitably leads to the need of discussing a taxonomy.
% Several strategies can be employed to group community discovery approaches into coherent families (e.g., considering the function they optimize, looking at the common constraints they impose on the identified community structures\dots): one that fits the dynamic context approached by DCD solutions relies on discussing their degree of \emph{smoothness}.

In the process of searching for communities over an evolving topology, one of the main questions that need to be answered is: how can the stability of the identified solution be ensured?
In static contexts, it has been shown that a generic algorithm executed on the same network that experienced a few topological variations - or even none in case of stochastic algorithms - might lead to different results \cite{aynaud2010static}.
The way Dynamic Community Discover (henceforth, DCD) algorithms take into account this problem plays a crucial role in the degree of stability of the solutions they can identify, i.e., on their \emph{smoothness}.
In \cite{rossetti2018community} DCD algorithms were grouped in three main categories, depending on the degree of smoothness they aim for:
\begin{itemize}
    \item Instant Optimal: it assumes that communities existing at time $t$ only depend on the current state of the network at $t$. Matching communities found at different steps might involve looking at communities found in previous steps, or considering all steps, but communities found at $t$ are considered optimal concerning the topology of the network at $t$. By definition, algorithms falling in this family are not temporally smoothed. Examples of Instant Optimal algorithms are \cite{Palla2007,rosvall2010mapping,Takaffoli2011,chen2010detecting}.
    
    \item Temporal Trade-off: it assumes that communities defined at time $t$ depend not only on the topology of the network at $t$ but also on the past topology, past identified partitions, or both. Communities at $t$ are therefore defined as a trade-off between an optimal solution at $t$ and the known past. They do not depend on future topological perturbations. Conversely, from Instant Optimal approaches, the Temporal Trade-off ones are incrementally temporally smoothed. Examples of Temporal Trade-off algorithms are \cite{gorke2010modularity,cazabet2010detection,rossetti2017tiles,folino2010multiobjective}.
    
    \item Cross-Time: algorithms of this class focus on searching communities relevant when considering the whole network evolution. Methods of this class search a single temporal partition that encompasses all the topological evolution of the observed network: communities identified at time $t$ depend on both past and future network structures. Methods in this class produce communities that are completely temporally smoothed. Examples of Cross-Time algorithms are \cite{aynaud2011multi,matias2017statistical,ghasemian2016detectability,jdidia2007communities,viard2016computing}.
\end{itemize}

All three classes of approaches have advantages and drawbacks; none is superior to the other since they model different DCD problem definition. 
Nevertheless, we can observe how each one of them is more suitable for some specific use cases.
For instance, if the final goal is to provide on-the-fly community detection on a network that will evolve in the future, Instant Optimal and Temporal Trade-off approaches represent the most suitable fit since they do not require to know in advance all the topological history of the analyzed network. 
Moreover, if the context requires working with a fine temporal granularity, therefore modeling the observed phenomena with link streams instead of snapshots, it is suggested to avoid methods of the first class, which are usually defined to handle well defined - stable - topologies.

Temporal smoothness and partition quality often play conflicting roles.
We can observe, for instance that, usually:
\begin{itemize}
    \item Instant Optimal approaches are the best choice when the final goal is to provide communities that are as good as possible at each step of the evolution of the network;
    \item Cross-Time approaches are the best choice when the final goal is to provide communities that are coherent in time, particularly over the long term;
    \item Temporal Trade-off approaches represent a trade-off between these other two classes: they are the best choice in the case of continuous monitoring, rapidly evolving data, and in some cases, limited memory applications.
    However, they can be subject to ``avalanche" effects due to the limited temporal information they leverage to identify communities (i.e., partitions evolve based on local temporal-optimal solutions that, on the long run may degenerate). 
\end{itemize}

\begin{figure*}[t]\centering
\centering
\includegraphics[width=0.82\textwidth]{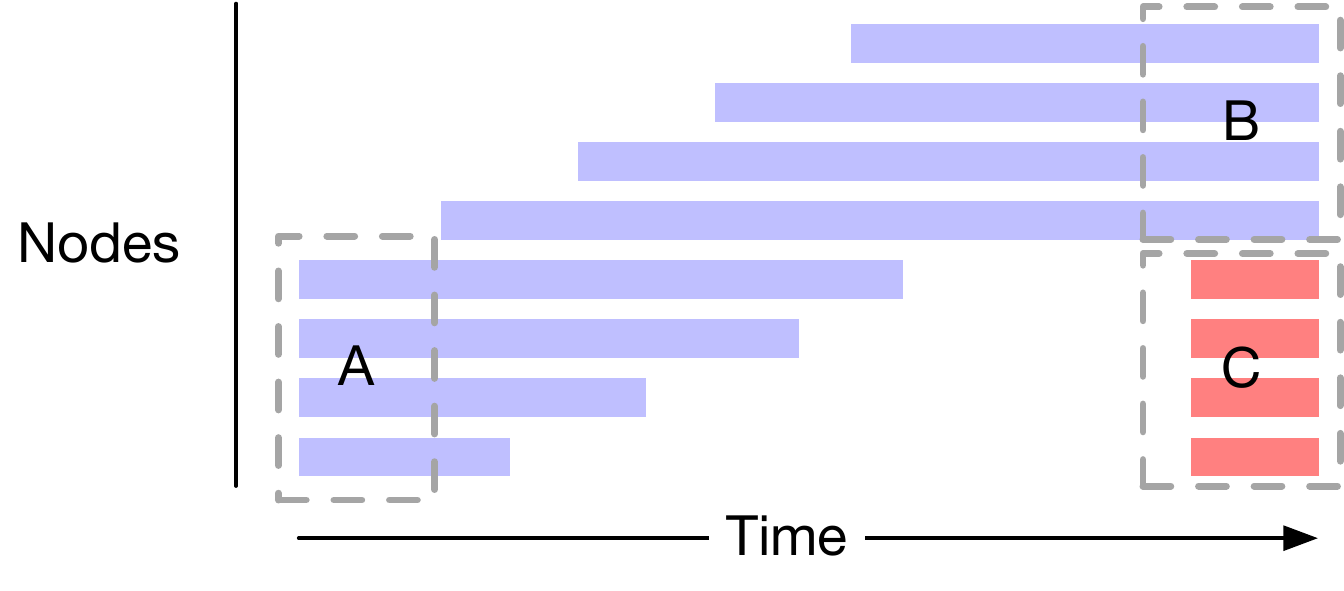}
\caption{Illustration of the ship of Theseus paradox. Each horizontal line represents a node. A same color represents nodes belonging to the same community according to a topological criterion (e.g., SBM). The community A is progressively modified until reaching state B. Community C is composed of the same nodes as the other community at its start. Which cluster (B or C) has the same identity as A? What if all details of the evolution are not known?}
\label{img:theseus}
\end{figure*}

\subsection{Preservation of identity: the ship of Theseus paradox}
The smoothness problem affects the way nodes are split into communities at each time. A different notion is the question of identity preservation along time, which arises in particular in case of a continued slow evolution of communities. It is well illustrated by the paradox of the \textit{ship of Theseus}. It is originally an ancient thought experiment introduced by Plutarch about the identity of an object evolving through time. It can be formulated as follows:
\vspace{5mm}

Let's consider a famous ship, the \textit{ship of Theseus}, composed of planks, and kept in a harbor as a historical artifact. As time passes, some planks deteriorate and need to be replaced by new ones. After a long enough period, all the original planks of the ship have been replaced. Can we consider the ship in the harbor to still be  \textit{the same} ship of Theseus? If not, at which point exactly did it ceased to be the same ship?

Another aspect of the problem arises if we add a second part to the story. Let's consider that the removed planks were stored in a warehouse, cleaned, and that a new ship, identical to the original one, is built with them. Should this ship, just built out, be considered as the \textit{real} ship of Theseus, because it is composed of the same elements? 
\vspace{5mm}

Let's call the original ship $A$, the ship that is in the harbor after all replacements $B$, and the reconstructed from original pieces, $C$. 
In terms of dynamic community detection, this scenario can be modeled by a slowly evolving community $c_1$ ($c_1= A$), from which nodes are removed one after the others, until all of them have been replaced ($c_1=B$). A new community $c_2$ appearing after that, composed of the same nodes as the original community ($c_2= C$). See fig. \ref{img:theseus} for an illustration.
A static algorithm analyzing the state of the network at every step would be able to discover that there is, at each step, a community ($c_1$, slowly evolving), and, at the end of the experiment, two communities ($c_1$ and $c_2$). But the whole point of dynamic community detection is to yield a longitudinal description, and therefore, to decide when two ships at different points in time are \textit{the same} or not.

This problem has barely been considered explicitly in the literature. However, each algorithm has implicitly to make a choice between which ship is the \textit{true} ship of Theseus. For instance, methods that are based on a successive match of communities, such as \cite{greene2010tracking}, consider that $A$ and $B$ are the same boats, but not $A$ and $C$. On the contrary, a method that matches similar clusters without the constraint of being consecutive, such as \cite{Falkowski2006}, consider that $C$ is more likely than $B$ to be the same ship than $A$.
Finally, methods such as \cite{Mucha2010} allow to set what is the influence of time on similarity, and therefore, to choose between those two extreme solutions.

\subsubsection*{Challenges}
The question of identity preservation in dynamic communities has been little discussed and experimented in the literature. For the sake of simplicity, most proposed methods use a mechanism of iterative matching or update of communities and therefore ignore the similarity between ships $A$ and $C$. However, this situation is probably very common in real networks, for instance, when confronted with seasonal or other cyclical patterns, where groups can disband and reform later. Developing new methods aware of the choice made in terms of identity preservation is, therefore, a challenge for the community.

\subsection{Scalability and computational complexity}
Early methods for community detection in static graphs had high computational complexity (e.g., \cite{girvan2002community}), thus were not scalable to large graphs. One part of the success of methods such as louvain\cite{blondel2008fast} or infomap\cite{rosvall2008maps} is that they can handle networks of thousands of nodes and millions of edges. 

Dynamic graphs represent a new challenge in terms of complexity. Among existing algorithms, we can distinguish different categories
\begin{itemize}
\item  Those whose complexity depends on the \textit{average size of the graph}
\item  Those whose complexity depends on the \textit{number of graph changes}.
\end{itemize}

Let's consider the example of a (large) graph composed of $n$ nodes and $m$ edges at time $t$, and which is evolving at the speed of $k$ changes every step, for $s$ steps. Algorithms in the first category, such as \textit{identify \& match} methods, needs to first compute communities at every step, thus their complexity is proportional to $s\mathcal{O}^{CD}(n,m)+(s-1)\mathcal{O}^{\to}(n)$ with $\mathcal{O}^{CD}(n,m)$ the complexity of the algorithm used at each step, and $\mathcal{O}^{\to}(n)$ the complexity of the matching process for communities found on the $n$ nodes.

Conversely, the complexity of an algorithm that update communities at each step such as \cite{cazabet2010detection} is roughly proportional (after the initial detection) to $s\mathcal{O}^{+=}(k)$, with $\mathcal{O}^{+=}(k)$ the complexity of updating the community structure according to $k$ changes. 
As a consequence, the first category is more efficient in situations where $k$ is large, and $n/m$ are small, while the second is more efficient when $n/m$ are large and $k$ small.
The complexity is not necessarily imposed by the adopted definition of community. For instance, algorithms proposed in \cite{Palla2007} and \cite{boudebza2018} yields rigorously the same dynamic communities, but they belong respectively to the first and second categories, as studied in \cite{boudebza2018}.

Another aspect to consider is parallelization. Although the computation of $\mathcal{O}^{CD}$ on many steps might seem expensive, this task can straightforwardly be processed in parallel. On the contrary, methods involving smoothing, or updating the structure in order, cannot be parallelized, as they need to know the communities at time $t$ to compute communities at time $t+1$. One must, therefore, consider the properties of a temporal network to know which method will or will not be computationally efficient on it.

\subsubsection*{Challenges}
The complexity of DCD algorithms has barely been explored and represents an important challenge to consider in future works. It is important to note that when dynamic networks are considered at a fine temporal resolution as in link streams, the number of edges (interactions) can be much larger than the number of nodes. For instance, in the SocioPatterns Primary School dataset \cite{sociopattrens2011}, more than 77 000 interactions are observed in a period spanning two days, despite having only 242 nodes. Algorithms developed for static algorithms use the sparsity of networks to improve their efficiency, but such an approach might be less rewarding in temporal networks. Analyzing the complexities of existing algorithms and developing new ones adapted to fine temporal resolution is, therefore, a challenge for researchers of the field.

\section{Handling different types of temporal networks}
\label{differentTypes}
Temporal networks can be modeled in different ways. Among the most common framework, we can cite:
\begin{itemize}
\item \textbf{Snapshot sequences}, in which the dynamic is represented as an ordered series of graphs
\item \textbf{Interval graphs} (or series of change) \cite{Holme2013}, in which intervals of time are associated with edges, and sometimes nodes
\item  \textbf{Link streams} \cite{Latapy2018}, in which edges are associated with a finite set of transient interaction times.
\end{itemize}

Each DCD algorithm is designed to work on a particular type of network representation. For instance, \textit{Identify \& Match} approaches consists of first identifying communities in each snapshot, and then matching similar communities across snapshots. Such a method is therefore designed to work (only) with snapshot sequences.
However, as it has been done in several articles, datasets can be transformed from one representation to the other, for instance by aggregating link streams into snapshots (e.g., \cite{Mucha2010}), or into interval graphs (e.g., \cite{cazabet2012}); thus the \textit{representation} of the dynamic graph does not necessarily limit our capacity to use a particular algorithm on a particular dataset.

We think however, that one aspect of the problem, related to representation, has not yet been considered in the literature. Methods working with snapshots and with interval graphs make the implicit assumptions that \textit{the graph any point in time is  well defined}, i.e., that each snapshot --or the graph defined by all nodes and edges present at any time $t$-- is not null, has a well-defined community structure, and is somewhat similar to neighboring snapshots. Said differently, those methods expect \textit{progressively evolving} graphs. To the best of our knowledge, this question has not been studied in the literature. A practitioner creating a snapshot sequence from a link stream using a too short sliding window (e.g., a window of one hour in a dataset of email exchanges) might obtain a well-formed dynamic graph on which an \textit{Identify \& Match} method can be applied, but the results would be inconsistent, as the community structure would not persist at such scales. The same dataset analyzed using longer sliding windows might provide insightful results. The problem is particularly pregnant for interval graphs, that can represent real situations of very different nature. For instance, an interval graph could represent relations (friend/follower relation in social networks) as well as interactions (phone calls, face-to-face interactions, etc.). It is clear that both networks should not be processed in the same way.

\subsubsection*{Challenges}
A challenge in the field will be to define the conditions of applicability of different methods better, and theoretical grounds to define when a network needs transformation to become suitable to be analyzed by a given method.

\section{Evaluation of Dynamic communities}

We have seen in previous sections that several approaches and methods exist to discover communities in temporal networks. In this section, we first discuss the evaluation of community quality. This process often requires the generation of dynamic networks with community structures, the topic of the second part of the section.

\subsection{Evaluation methods and scores}
As already discussed, there is not a single, universal definition of what is a good community and, consequently, no unique and universal way to evaluate their quality. Nevertheless, for static communities, many functions have been proposed, to evaluate them either i) intrinsically (internal evaluation), by means of quality functions, (e.g., Modularity, Conductance, etc.) and ii)Relatively to a reference partition (external evaluation), using a similarity function (e.g., NMI, aNMI, etc.). Both approaches have pros and cons that have been thoroughly discussed in the literature \cite{peel2017ground,yang2015defining}. Few works have been done to extend those functions to the dynamic case. 

\subsubsection{Internal evaluation}
In most works, static quality functions are optimized at each step, often adding a trade-off of similarity with temporally adjacent partitions to improve community smoothness (see section \ref{smoothness}).
Some works are based on a longitudinal adaptation of the modularity \cite{Mucha2010,aynaud2010static}, but they require to create a new graph with added inter-snapshot edges, and therefore cannot be used to evaluate algorithms based on different principles. Works based on Stochastic Block Model \cite{matias2017statistical,Yang2009} also optimize a custom longitudinal quality function.

\subsubsection{External evaluation}
Articles doing external evaluation requires to have a reference partition. Since few annotated datasets exist, a synthetic generator is used (see section \ref{generator}. 
The comparison often uses the average of a static measure (e.g., NMI) computed at each temporal step \cite{bazzi2016generative}, eventually weighted to take into account the evolution of network properties \cite{rossetti2017}.
A notable exception is found in \cite{granell2015benchmark}, where windowed versions of similarity functions (Jaccard, NMI, NVI) are introduced, by computing their contingency table on two successive snapshots at the same time.

\vspace{5mm}
\subsubsection*{Challenges}
The evaluation of the quality of dynamic communities, both internally and externally, certainly represents a challenge for future works in dynamic community detection. Methods directly adapted from the static case do not consider the specificity of dynamic communities, in particular, the problems of smoothness and community events. This question is of utmost importance, since, despite the large variety of methods already proposed, their performances on real networks besides the ones they have been designed to work on is still mostly unknown.

\subsection{Generating dynamic graphs with communities}
\label{generator}
%Several network properties can be used to characterize real-world phenomena: network modeling aims to replicate them, thus allowing for the generation of synthetic datasets that, at least to some extent, can be used as analytical proxies. 
%The general aim of network modeling is to capture some essential properties lying behind real-world phenomena and replicate them while generating synthetic data, imposing only a few simple constraints.
Complex network modeling studies gave birth to a new field of research: synthetic network generators. 
Generators allow scientists to evaluate their algorithms on synthetic data whose characteristics resemble the ones that can be observed in real-world networks. The main reason behind the adoption of network generators while analyzing the performance of a dynamic community detection (DCD) algorithm is the ability to produce benchmark datasets that enable i) Controlled environment testing, e.g., in term on network size, dynamics, structural properties, etc., and ii) comparison with a planted ground-truth.
% \begin{itemize}
%     \item Controlled environment testing: Network generators allow fine tuning of network characteristics, such as network size and density. Such flexibility enables an extensive algorithm evaluation on networks that have different characteristics but are generated to follow similar topologies. 
%     Generators make it possible, given a CD algorithm, to evaluate:
%     \begin{itemize}
%         \item Stability: The performance of a CD approach can be evaluated on a high number of network instances with similar properties to provide an estimate of the algorithmic stability.
%         \item Scalability: Synthetic graphs can be used to test the actual scalability of an algorithm while increasing the network size.
%     \end{itemize}
%     \item Ground-truth testing: Some network generators provide as a by-product a ground-truth partition of the generated network. Such a partition can be used to evaluate the one provided by the tested algorithm.
% \end{itemize}

Two families of network generators have been described to provide benchmarks for DCD algorithms: generators that produce static graphs-partitions and generators that describe dynamic graphs-partitions. 
%Surprisingly, the former are often used to evaluate DCD algorithms: this happens because they are more widespread than the latter and allows comparison with static algorithms. 
Static graphs are used to evaluate the quality of the detection at a single time $t$, and cannot inform about the smoothness of communities. The most known are the GN benchmark \cite{girvan2002community}, the LFR benchmark \cite{lancichinetti2009benchmarks} and planted partitions according to the stochastic block model.

Several methods have been proposed to generate dynamic networks with communities. The network can be composed of a sequence of snapshots, as in \cite{bazzi2016generative}, in which, at each step, the community structure (based on an SBM) drifts according to a user-defined inter-layer dependency.
Another approach consists in having an initial partition yielded by a static algorithm (LFR in \cite{greene2010tracking}, GN in \cite{lin2008}), and to make it evolves randomly \cite{greene2010tracking} or until reaching an objective network with a different community structure\cite{lin2008}.

Finally, another class of methods generates slowly evolving networks whose changes are driven by community events --merge, split, etc.-- that can be tuned with parameters such as the probability of event occurrences. One of these methods is RDyn \cite{rossetti2017}, whose communities are based on a similar principle than LFR. Another method has been proposed in Sengupta et al. \cite{sengupta2017benchmark}, which has the particularity of generating overlapping community structures.

\subsubsection*{Challenges}
As we have seen, various methods already exist to generate dynamic graphs with slowly evolving communities. They have different properties, such as community events, stable edges, or overlapping communities. Active challenges are still open in this domain, among them i)The generation of link streams with community structures, ii)The empirical comparison of various DCD methods on those benchmarks, and iii)An assessment on the realism of communities generated with such benchmarks, compared with how empirical dynamic communities behave.

\section{Libraries and standard formats to work with dynamic communities}
In recent years, many tools and software have been developed to manipulate and process network data. Many of those tools have implemented community detection algorithms. Among the best known, we can cite networkx \cite{hagberg2008exploring}, iGraph \cite{igraph} and snap \cite{leskovec2016snap}, which propose a wide variety of network analysis tools, among them community detection algorithms, and related quality functions and scores. Some libraries are even designed specifically for community detection such as CDlib\footnote{\url{https://github.com/GiulioRossetti/cdlib/tree/master/cdlib}}. However, none of them can deal with dynamic networks. Very recently, a few libraries have been introduced to work with dynamic networks, such as tacoma\footnote{\url{https://github.com/benmaier/tacoma}} and pathpy \cite{scholtes2017network} but do not include community detection algorithms. 

Furthermore, no standard format has yet emerged to represent dynamic communities and their evolution, which is particularly a problem to compare solutions yielded by different methods. 
This lack of common tools and standard representation certainly represents an obstacle, and a challenge to overcome for the DCD research community.

\section{Conclusion}
In this chapter, we have introduced the theoretical aspects of dynamic community detection and highlighted some of the most interesting challenges in the field. Among them, we think that a better formalism to represent the evolution of dynamic clusters and their events, in particular in the context of gradually evolving communities, would facilitate the comparison and the evaluation of communities and detection methods. The scalability of existing approaches is also a concern, again, in the context of \textit{link streams} or other temporal networks studied at fine temporal scales. Finally, a recently introduced technique \textit{graph embedding}, has attracted a lot of attention in various domains. Applications exist to temporal networks, although no work has focused on the dynamic community detection problem yet, to the best of our knowledge. Using this new technique to propose scalable methods could be another challenge worthy of investigation.
\bibliographystyle{unsrt}
\bibliography{ref.bib}

\end{document}